\documentclass{ceurart}

\usepackage{listings}
\usepackage{subcaption}
\usepackage[inline]{enumitem}
\usepackage{tabularx}
\begin{document}

\copyrightyear{2022}
\copyrightclause{Copyright for this paper by its authors.
  Use permitted under Creative Commons License Attribution 4.0
  International (CC BY 4.0).}

\conference{CREAI 2022 - Workshop on Artificial Intelligence and Creativity,
  November 28-December 2, 2022, Udine, Italy}

\title{Pitchclass2vec: Symbolic Music Structure Segmentation with Chord Embeddings}


\author[1]{Nicolas Lazzari}[%
orcid=0000-0002-1601-7689,
email=nicolas.lazzari2@studio.unibo.it,
]
\cormark[1]
\fnmark[1]
\address[1]{Department of Computer Science and Engineering, University of Bologna, Mura Anteo Zamboni, 7, Bologna 40126, Italy}

\author[1]{Andrea Poltronieri}[%
orcid=0000-0003-3848-7574,
email=andrea.poltronieri2@unibo.it,
]
\fnmark[1]

\author[2]{Valentina Presutti}[%
orcid=0000-0002-9380-5160,
email=valentina.presutti@unibo.it,
]
\fnmark[1]
\address[2]{LILEC, University of Bologna, Via Cartoleria, 5, Bologna 40124, Italy}

\cortext[1]{Corresponding author.}
\fntext[1]{Alphabetical order.}

\begin{abstract}
Structure perception is a fundamental aspect of music cognition in humans. Historically, the hierarchical organization of music into structures served as a narrative device for conveying meaning, creating expectancy, and evoking emotions in the listener.
Thereby, musical structures play an essential role in music composition, as they shape the musical discourse through which the composer organises his ideas.
In this paper, we present a novel music segmentation method, pitchclass2vec, based on symbolic chord annotations, which are embedded into continuous vector representations using both natural language processing techniques and custom-made encodings. Our algorithm is based on long-short term memory (LSTM) neural network and outperforms the state-of-the-art techniques based on symbolic chord annotations in the field.
\end{abstract}


\maketitle

\section{Introduction}
\label{sec:introduction}

One of the main factors that influence music perception is the hierarchical structure of music compositions. 
Regardless of their level of musical knowledge and harmonic sensitivity \cite{tan1981harmonic} or their cultural origins \cite{stevens2012music}, listeners are able to use intuitive knowledge to organize their perception of musical structures \cite{chiappe1997phrasing}. 
Indeed, there is empirical evidence that neural activity correlates with musical structure in listeners' perception \cite{knosche2005perception}. The structuring and predictability of musical compositions is also recognised as a viable therapy in the treatment and assessment of children and adolescents with autistic spectrum disorder \cite{wigram2006music}.

Music structuring is one of the tools used by composers to tell a story. According to \cite{burns1987typology} ``Music-making is, to a large degree, the manipulation of structural elements through the use of repetition and change.''. The repetition of harmonic progressions (sequences of chords), in particular in the context of western tonal music, gives to artists the ability to guide listeners through a journey that creates dramatic narratives, conveying a sense of conflict that demands a solution \cite{temperley2004cognition}.

\medskip
\begin{figure}[ht]
    \centering
    \includegraphics[width=\textwidth]{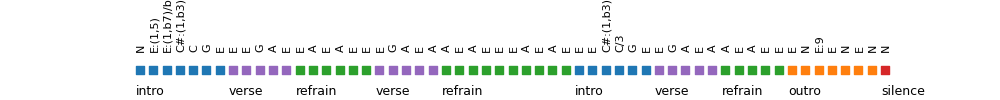}
    \caption{Structure of Helter Skelter by The Beatles. Chords are presented in Harte format \cite{harte2005symbolic}.}
    \label{fig:helterskelter-structure}
\end{figure}

Figure \ref{fig:helterskelter-structure} shows the structure of \textit{Helter Skelter} by The Beatles, highlighting the musical chords of the song.
In the example, by means of the alternation between \textit{verse} and \textit{refrain} the artist establishes a common repetitive pattern. The addition of an instrumental section after the second \textit{refrain} and the repetition of the \textit{intro} reinforces the repetitive aspect of the composition. The upcoming \textit{outro} section denies the expectation of a new \textit{verse}, right before the song ends. 
Expectation and the way it is fulfilled or denied is an essential part in musical enjoyment \cite{temperley2004cognition}.
In fact, it has been shown empirically that the emotional response to a musical composition varies as the degree of repetition changes \cite{livingstone2012emotional}. 
Understanding musical structures is hence fundamental in music analysis and composition. 
Artists can benefit from the feedback provided by a system able to highlight possible hierarchical structures in their compositions.

Music structure segmentation is a broad term related to the study of musical form, which describes how musical pieces are structured. In particular it can be divided in two main categories: phrase-structure segmentation and global segmentation \cite{giraud2016computational}. Phrase-structure consists in detecting sections from the melodic information of a piece. While the aim of phrase-structure is not to obtain a global segmentation, the detected sections provide valuable insights in the task of global segmentation.
In the following, we will refer to music structure segmentation as the task of global segmentation. Music structure segmentation is a music information retrieval (MIR) task that consists in identifying and labelling key music segments (e.g. \textit{chorus}, \textit{verse}, \textit{bridge}) of a music piece \cite{McCallum19unsupervised}. 
Given a musical composition, its musical segmentation consists in the identification of non-overlapping segments, which we will refer to as sections. Each section is characterized by a label that classifies its function such as \textit{intro} or \textit{verse} in figure \ref{fig:helterskelter-structure}. 
A correct segmentation does not necessarily assign the correct labels to each section of the composition, but rather focuses on the correct estimation of the boundaries of each section. 
Once boundaries has been accurately predicted, a labeling process is performed to obtain the final annotation \cite{nieto2020audio}.

Most of the recent methods and research approaches are based on audio analysis techniques \cite{nieto2020audio}, nonetheless harmonic information, isolated from tempo and rhythm, have been successfully used in several tasks in the field of music information retrieval (MIR) (e.g. \cite{de2013geometrical, de2013structural}).

In this paper, we focus on the music structure segmentation task by only taking into account harmonic information extracted from symbolic notations (music chord annotations). 
The assumption behind this approach is that the identification of harmonic sub-sequences (harmonic patterns) can be influential in defining the structure of a song and the sections of which it is composed.
For instance, by taking a closer look at Figure \ref{fig:helterskelter-structure} it is easy to notice how harmonic information can provide valuable information in the structure segmentation task: all \textit{verse}s are roughly based on the same harmonic progression (\textit{E}, \textit{G}, \textit{A}, \textit{E}) while \textit{refrain}s are based on a different harmonic progression (\textit{A}, \textit{E}, \textit{A}, \textit{E}, \textit{E}). 
A segmentation strongly based on those recurrent patterns is likely to be coherent with the way the composer shaped the progression in the first place.

\medskip
The objective of this paper is threefold: 
\begin{enumerate*}[label=(\roman*)]
    \item we propose \textit{pitchclass2vec}, a novel chord embedding method;
    \item we use this encoding with a recurrent neural network on a corpus of musical chords; and
    \item we compare the performance of the encoding with the state-of-the art methods in the field.
\end{enumerate*}

The chord embedding method proposed, \textit{pitchclass2vec}, encodes a chord using a one-hot encoding of the notes that compose it by making use of word embedding techniques. 
Each embedded chord is defined to be similar to the embedding of its neighbouring chords in an harmonic progression. This formalization is supposed to approximate the semantic meaning of a chord \cite{sahlgren2008distributional} and has been widely used in the natural language processing field \cite{mikolov2013word2vec, bojanowski2018fasttext}.
We use \textit{pitchclass2vec} embeddings to train an LSTM neural network that predicts the section of each chord. 
Through its recurrent layers the neural network is able to learn relationships between the elements of a sequence. 
This allows the model to detect repetitive patterns of the harmonic progression and predict the a segmentation of the whole composition.  The model provides a baseline to test the efficacy of the proposed chord embedding method.
State-of-the-art results are achieved in the task of music structure segmentation on symbolic harmonic data, providing evidences that \textit{pitchclass2vec} is able to provide accurate chord representations.
However, the embedding method employed here for the music segmentation task, can be used in a variety of applications in the field of Music Information Retrieval, such as retrieving harmonically related pieces \cite{de2013geometrical}, automatic chord recognition \cite{ohanlon2021fifthnet} and music genre classification \cite{liang2020pirhdy}.


\medskip
The paper is organised as follows: Section \ref{sec:related} introduces the related works, Section \ref{sec:model} describes the novel chord embedding method and the recurrent neural network used for the segmentation task. Section \ref{sec:experiments} presents the experiments performed and Section \ref{sec:results} gives an overview of the obtained results. Finally in Section \ref{sec:conclusion} we discuss the results and new research directions to be explored.

\section{Music segmentation: state of the art}
\label{sec:related}

Automatic segmentation on audio signal is a prolific research field in which many different solutions have been presented, ranging from self-similarity matrices \cite{weiss2011unsupervised, bello2010identifying} to neural network based methods \cite{shibata2020music, wang2021supervised, marmoret2021uncovering}. Harmonic content has been used to improve those methods both using probabilistic models \cite{pauwels2013combining} and transformer based models \cite{chen2019harmony}.
Significant research has been performed on phrase-level structural segmentation based on melodic \cite{frankland2004structuralMelody, cambouropoulos2001local, velarde2013approach} as well as polyphonic content \cite{meredith2002algorithms}.
However, to the best of our knowledge, the only approach proposed in literature for global music segmentation on symbolic harmonic content is FORM \cite{de2013structural}.

FORM performs structural segmentation by exploiting repeated patterns extracted from harmonic progressions encoded as sequences of strings. Each string represent a chord. In the original work chord labels are transformed into $24$ class of chords, $12$ major chords and $12$ minor chords, while every other chord feature is removed. 
In this paper, FORM is re-implemented in order to compare the results of the proposed method with the current state of the art (see Section \ref{sec:results}).
FORM pattern detection algorithm is based on suffix trees. Each node on a suffix tree represents a (possibly recurrent) sub-sequence of a string. FORM extracts sub-sequences appearing in at least two position in the analyzed harmonic progression.
A partial segmentation is obtained by labeling each sub-sequence as a new section. The final segmentation is obtained by labeling remaining sub-sequences as their preceding neighbouring section.
The results are compared with a random baseline that generates arbitrarily long structures and to a heuristic that assigns to each composition the typical pop song structure \textit{ABBBBCCBBBBCCDCCE} \cite{de2013structural}, in which each different label represent a structure in the chord progression and is stretched to fit the whole sequence.
The main issue with FORM is in way chord labels are compared. The string representation does not take into account semantic similarity between chords nor the algorithm is able to detect near-similar patterns, i.e. patterns whose difference can be ignored in the context of music structure segmentation.

\medskip
Our structure segmentation method is based on our novel chord representation method, \textit{pitchclass2vec}, based on continuous word representation.
The core idea of continuous word representation is based on the Distributional Hyphotesis \cite{sahlgren2008distributional}: the semantic meaning of a word $w$ can be approximated from the distribution of words that appear within the context of $w$.
The objective of continuous word representation is the maximization of the following log-likelihood:
\begin{displaymath}
  \sum_{t=1}^{T} \sum_{c \in \mathcal{C}_t} \log p(w_c | w_t),
\end{displaymath}
where $\mathcal{C}_t$ represents the indices of the words that appears as context of the word $w_t$ and the function $p(w_c | w_t)$ is parameterized using $d$-dimensional vectors in $\mathbb{R}^d$, respectively $\mathbf{u}_{w_t}$ and $\mathbf{v}_{w_c}$.
The problem can be framed as a binary classification task in which words are predicted to be present (or absent) from the context of $w_t$.
A similarity function $s(w_c, w_t)$ between two words $w_c$ and $w_t$, can be computed as the scalar product $\mathbf{u}_{w_t}^T \mathbf{v}_{w_c}$.
The representations obtained by training the described method on a large corpus correctly approximates the semantic meaning of words. In the last few years, continuous word representation has been applied in a growing number of application areas, achieving state-of-the-art results in the natural language processing field \cite{qilu2021briefsurvey} in tasks such as part of speech tagging\cite{meftah2018pos}, named entity recognition\cite{chiu2016ner} and document classification\cite{lilleberg2015support}.

\medskip
\begin{figure}
\centering
\begin{subfigure}{.5\textwidth}
  \centering
  \includegraphics[height=14em,keepaspectratio]{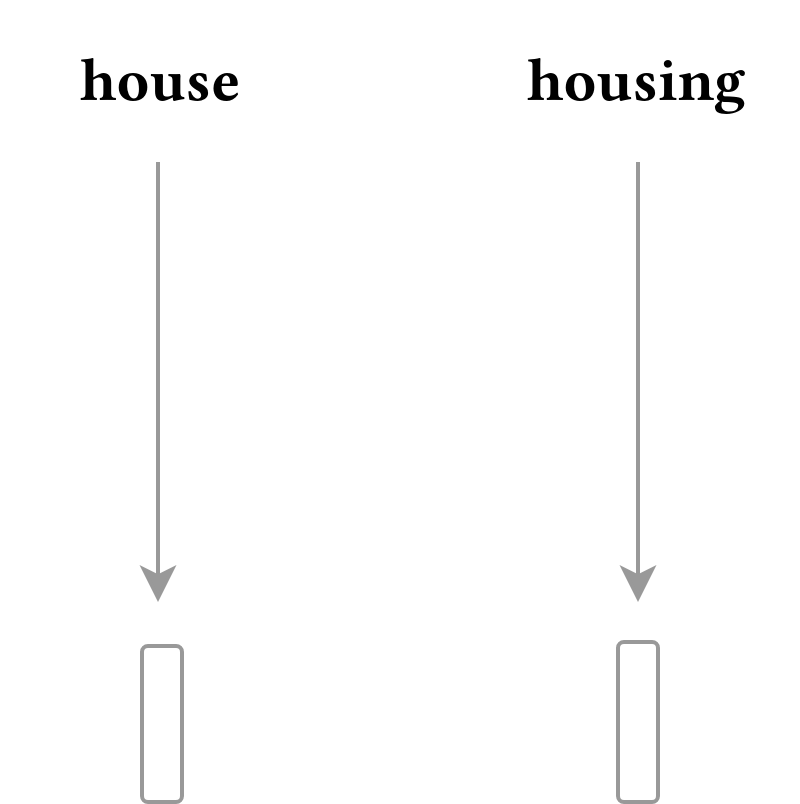}
  \caption{word2vec}
  \label{fig:word2vec}
\end{subfigure}%
\begin{subfigure}{.5\textwidth}
  \centering
  \includegraphics[height=14em,keepaspectratio]{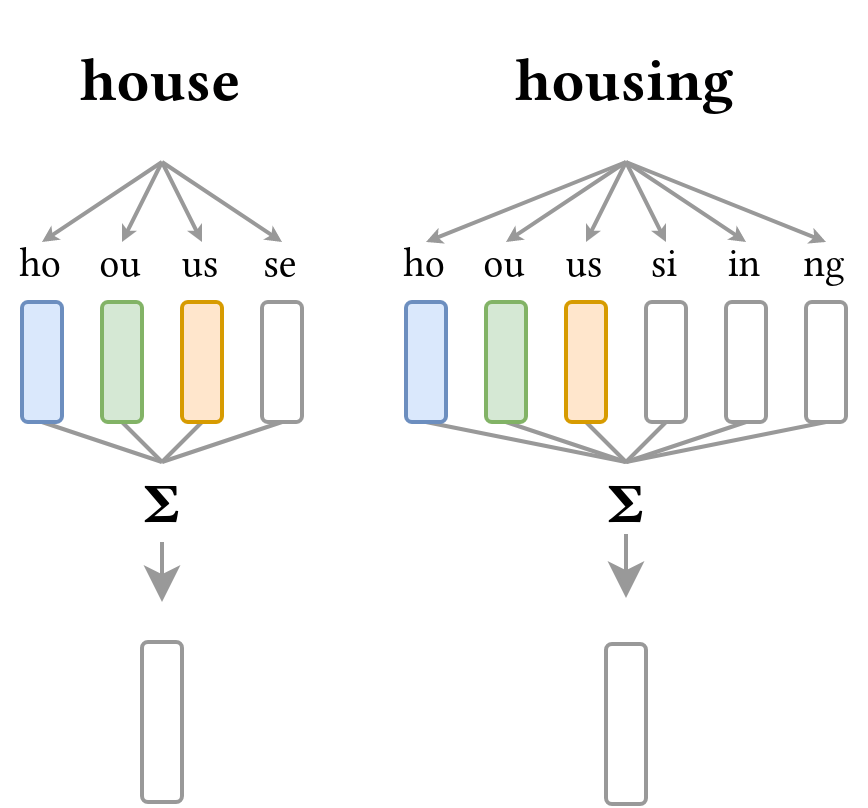}
  \caption{fasttext}
  \label{fig:fasttext}
\end{subfigure}
\caption{ \textit{Word2vec} (a) and \textit{fasttext} (b) embedding methods. With \textit{word2vec} each embedding is computed independently of its morphological structure. \textit{Fasttext} instead compute the representation as the sum of the n-grams that compose a word. Words that share one or more n-grams have a similar representation as they are computed in a similar way. In the example 2-grams are represented but in general n-grams up to the length of the term are commonly used.}
\label{fig:word2vec-vs-fasttext}
\end{figure}

The described approach has been first proposed by the \textit{word2vec} skipgram model \cite{mikolov2013word2vec}. \textit{Word2vec}, however, is limited by the lack of morphological knowledge of a word. 
When computing the representation of a word, none of its morphological components are taken into account. Let's take for instance two morphologically similar words, \textit{house} and \textit{housing}. Their computation does not share any common element and the final representation of the words will not be influenced by their similarities.
\textit{Fasttext} \cite{bojanowski2018fasttext} was presented as a solution to this issue and has proven to be more effective in the representation of a word using continuous representations. 
The novel aspect is in the way representations are computed. At first the n-grams that compose a word are extracted. For each n-gram a continuous vector representation is computed, using the same methodology as \textit{word2vec}. The representation of the original words is finally obtained as the sum of its n-gram components. 
Using this technique, the final representation of a word is conditioned by its morphological structure. When two words share one or more n-grams their vectors will be the sum of at least one common element, which will bias both vectors in being more similar to each other. 
An additionally advantage of the \textit{fasttext} approach is the way out-of-vocabulary words (words that never appear in the training corpus, and whose representation is hence unknown) are handled. When using fixed approach such as \textit{word2vec}, out of vocabulary words are represented as a static vector, usually randomly sampled from a normal distribution. \textit{Fasttext} instead is able to compute the representation in a meaningful way, given that at least one of the n-grams in the out-of-vocabulary term has been computed previously in the training corpus. 
Figure \ref{fig:word2vec-vs-fasttext} shows a visual comparison between \textit{word2vec} (Figure \ref{fig:word2vec}) and \textit{fasttext} (Figure \ref{fig:fasttext}).

Continuous word representations have already been applied to chord symbols with promising results. In \textit{chord2vec} \cite{madjiheurem2016chord2vec} the authors obtain state-of-the-art results on the log-likelihood estimation task. Log-likelihood estimation is the task of correctly estimating, given one element in a sequence, the probabilities of another element being the upcoming element in the sequence. 
\textit{Chord2vec} is inspired by the \textit{word2vec} method, in which chords are represented by the notes that they are composed of. 
The representation model proposed by \textit{chord2vec} is similar to \textit{pitchclass2vec}. 
Instead of computing chord representations with the notes that compose a chord, the representations of \textit{pitchclass2vec} takes into account the relationship between the notes that compose each chord. An in depth discussion is presented in section \ref{sec:model}.

More recently, \textit{word2vec}-based approach on symbolic chord annotations has been analyzed by \cite{anzuoni2021historical}. 
Chord representations are based on the chord label without taking into account the notes that compose it. The encoding is then used on two different tasks: chord clustering and log-likelihood estimation. 
The log-likelihood estimation task is used to investigate the historical harmonic style of different composers. 
The log-likelihood results strongly correlates with current musicological knowledge. For instance the model finds difficult to predict chords from artists that make a sporadic use of common harmonic progressions \cite{anzuoni2021historical}. 
FThe chord clustering task highlights how it's possible to observe similarities between \textit{functionally equivalent} chords (chords that shares notes between each other) and a well defined difference between \textit{functionally different} chords. 
Continuous word representations are hence adequate to encode chords in the first place, and more importantly they are able to autonomously internalize relations between chords that have been previously observed by domain experts.

\section{Pitchclass2vec model}\label{sec:model}

Embedding approaches used in natural language processing have obvious limitations when it comes to dealing with musical content, such as musical chords.
While relying on purely syntactical representations has been show to correctly encapsulate some forms of domain knowledge \cite{anzuoni2021historical}, more advanced representations are needed to obtain accurate results when dealing with harmonic progressions \cite{madjiheurem2016chord2vec}.

There are, however, some ambiguous cases in which both vector representations might introduce wrong similarities between chords.
Let us take for instance the chords \textit{C:maj} and \textit{C:maj13}\footnote{In Harte\cite{harte2005symbolic} notation}, whose notes are respectively $\mathcal{C}_{\textit{C:maj}} = \{C, E, G\}$ and $\mathcal{C}_{\textit{C:maj13}} = \{ C, E, G, B, D, A \}$. Both chords' labels only differ by two characters, however the difference between the notes that they are composed of can't be neglected. A method exclusively based on syntactical information would wrongly represent the vectors as similar between each other.
Conversely, only relying on the notes that compose a chord results in ambiguous representations of some particular classes of chords, called \textit{enharmonic} chords.
For instance, the \textit{enharmonic} chords \textit{C:dim} and \textit{Eb:dim} share the exact same set of notes, $\mathcal{C}_{\textit{C:dim}} = \mathcal{C}_{\textit{Eb:dim}} = \{C, Eb, Gb, A\}$ but need to be represented as different chords as they serve different harmonic purposes. \textit{Chord2vec} would wrongly represents both chords as the same exact vector.

In order to overcome the aforementioned limitations, we propose an encoding which requirements can be summarised as follows:
\begin{enumerate*}
    \item it has to be based on the constituent notes of a chord, rather than its label; and
    \item it must take into account the relation between those notes instead of the notes themselves.
\end{enumerate*}

The proposed encoding is grounded on tonal music theory: each chord $c$ is composed of a set of notes $\mathcal{C} \subset \mathcal{N}$, where $\mathcal{N}$ is the set of all notes and $C$ is called the \textit{pitch class} of a chord. An important distinction is represented by the \textit{root} note, which names the chord and plays an important role in its harmonic function.
 
\medskip
\begin{figure}
\centering
\begin{subfigure}{.48\textwidth}
  \centering
  \includegraphics[width=\textwidth]{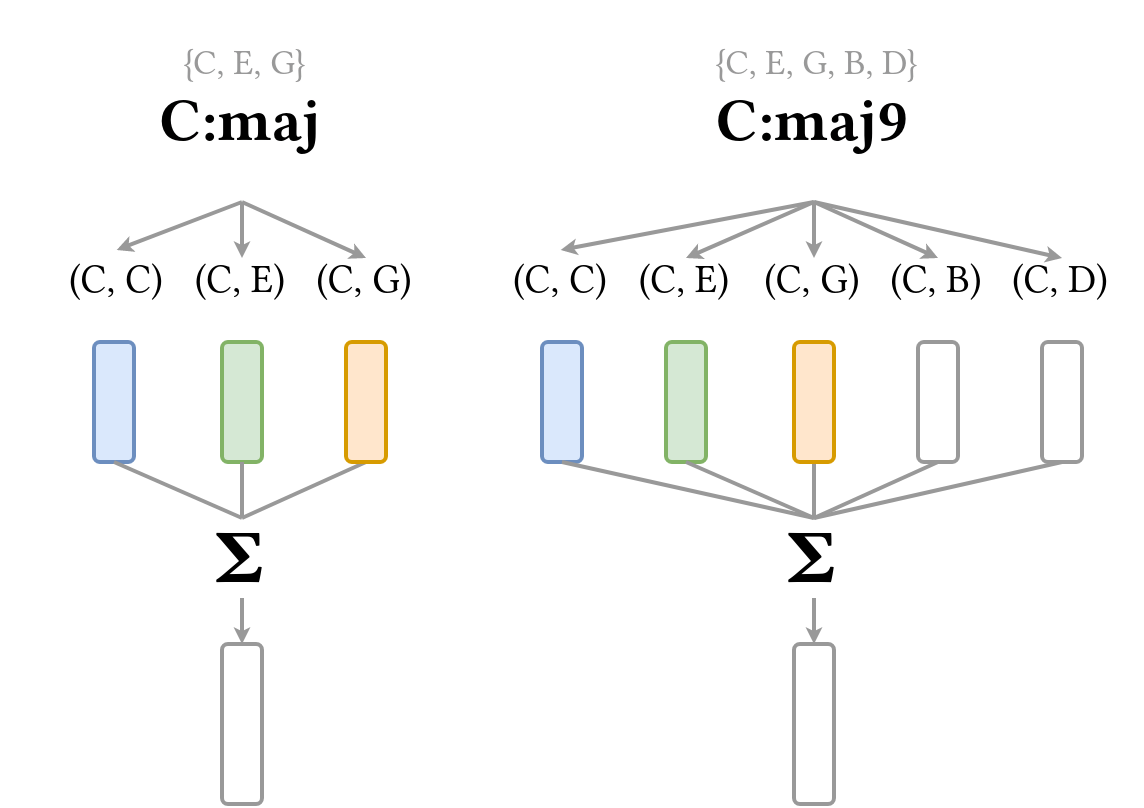}
  \caption{C:maj and C:maj9 chord embeddings. The final representation is computed from common elements and will hence share some aspects.}
  \label{fig:pitchclass2vec-similarlabel-embedding}
\end{subfigure}
\hfill
\begin{subfigure}{.48\textwidth}
  \centering
  \includegraphics[width=\textwidth]{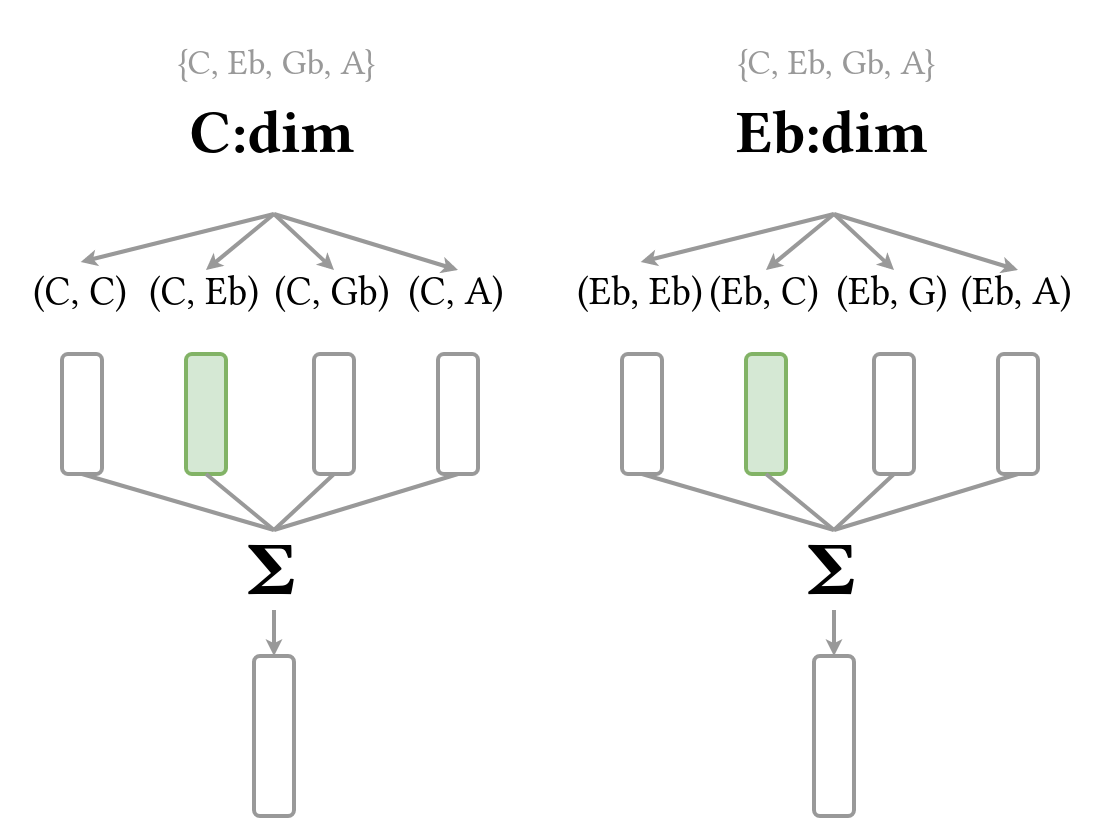}
  \caption{C:dim and Eb:dim chord embeddings. Both chords are composed of the same notes but using mostly different components.}
  \label{fig:pitchclass2vec-enharmonic-embedding}
\end{subfigure}
\caption{Visual reference on pitchclass2vec embedding method.}
\end{figure}

We encode each chord as the Cartesian product $\mathcal{I}_c = \textit{root}_c \times \mathcal{C}_c$ between the \textit{root} note and the \textit{pitch class} of the chord. The vector representation $\mathbf{u}_c$ of a chord $c$ is computed as
\begin{displaymath}
  \mathbf{u}_c = \sum_{i \in \mathcal{I}_c} \mathbf{u}_i
\end{displaymath}
where $\mathbf{u}_i$ is the vector representation of the tuple $x_i \in \mathcal{I}_c$. 
See Figure \ref{fig:pitchclass2vec-enharmonic-embedding} for a visual reference on how \textit{pitchclass2vec} handles \textit{enharmonic} chords and Figure \ref{fig:pitchclass2vec-similarlabel-embedding} on how chords with common components are handled.
This formalization can be seen as an extension of the chord2vec \cite{madjiheurem2016chord2vec} method, in which the chord inner structure is taken into consideration as well.


Nevertheless, the label of a chord has a well-defined semantic. Chords composed of the same set of notes may have different harmonic functions. For example, the chords \textit{G:min7} and \textit{Bb:6}, despite different labels contain the exact same notes: $\mathcal{C}_{\textit{G:min7}} = \mathcal{C}_{\textit{Bb:6}} = \{G, Bb, D, F\}$. 
This problem is particularly evident in datasets containing annotations made by experts, where the choice of label is the result of a meticulous analysis. 
For this reason, we have implemented two different variants of \textit{pitchclass2vec}: 
\begin{enumerate*}[label=(\roman*)]
    \item a variant combining the approach proposed by \textit{word2vec} with \textit{pitchclass2vec}; and 
    \item a variant combining \textit{fasttext} with \textit{pitchclass2vec}.
\end{enumerate*}

In order to obtain mixed embeddings we test different hybrid combinations before passing the new representation to the LSTM model:
\begin{enumerate}[label=(\roman*)]
    \item concatenating the embeddings;
    \item concatenating the embeddings and projecting the result in a $N$-dimensional vector, using a fully connected layer;
    \item projecting the embeddings in the same $N$-dimensional space by using two different fully connected layer and summing the $N$-dimensional vectors;
    \item computing a new representation of each embedding by using two separate LSTM layers and summing the resulting vectors;
    \item computing a new representation of each embedding by using two separate LSTM layers and concatenating the resulting vectors.
\end{enumerate}
None of the combination used proved to be able to outperform the others and we decided to stick to the first simpler and faster approach.

\subsection{Implementation details}
\label{sec:implementation-details}
The model is implemented using \texttt{pytorch}. We train the model on a set of $\approx 16000$ chord progressions (with a total number of over $1M$ chord instances), taken from the Chord Corpus (ChoCo) dataset \cite{deberardinis2022choco}. ChoCo is a chord dataset consistsing of more than $20000$ tracks taken from $18$ different professionally curated datasets. 
All datasets have been parsed in JAMS \cite{humphrey2014jams} format and converted in Harte Notation \cite{harte2005symbolic}.
We train the model for at most $10$ epochs on an \textit{NVIDIA RTX 3090} with batch sizes of $512$ harmonic progressions. 
We manually tune the batch size to efficiently train the model on our available resources.
For each chord we take a window of $4$ context chords as positive examples, $2$ preceding and $2$ succeeding, as it has been done in the original \emph{fasttext} implementation \cite{bojanowski2018fasttext}. Then, we sample $20$ random chords as negative examples. Even though it has been shown that windows of different sizes yields different results depending on the task they are applied to \cite{caselles2018word2vec} here we will rely on a fixed size window to better compare it to the related works.
We subsample our corpus to obtain a more balanced one by removing some of the most frequent chords instances.  We use a factor of $t = 10^{-5}$ as suggested by \cite{mikolov2013word2vec} to allow a faster and more accurate training phase.
The model is trained using a standard training procedure where a binary cross entropy loss between a chord and its positive and negative examples is minimized using Adam optimizer, with fixed learning rate of $0.025$.
We set the embedding dimension to $10$ as the result of manual trials.

\section{Experimental setup}\label{sec:experiments}
\begin{figure}
    \centering
    \includegraphics[width=.7\textwidth]{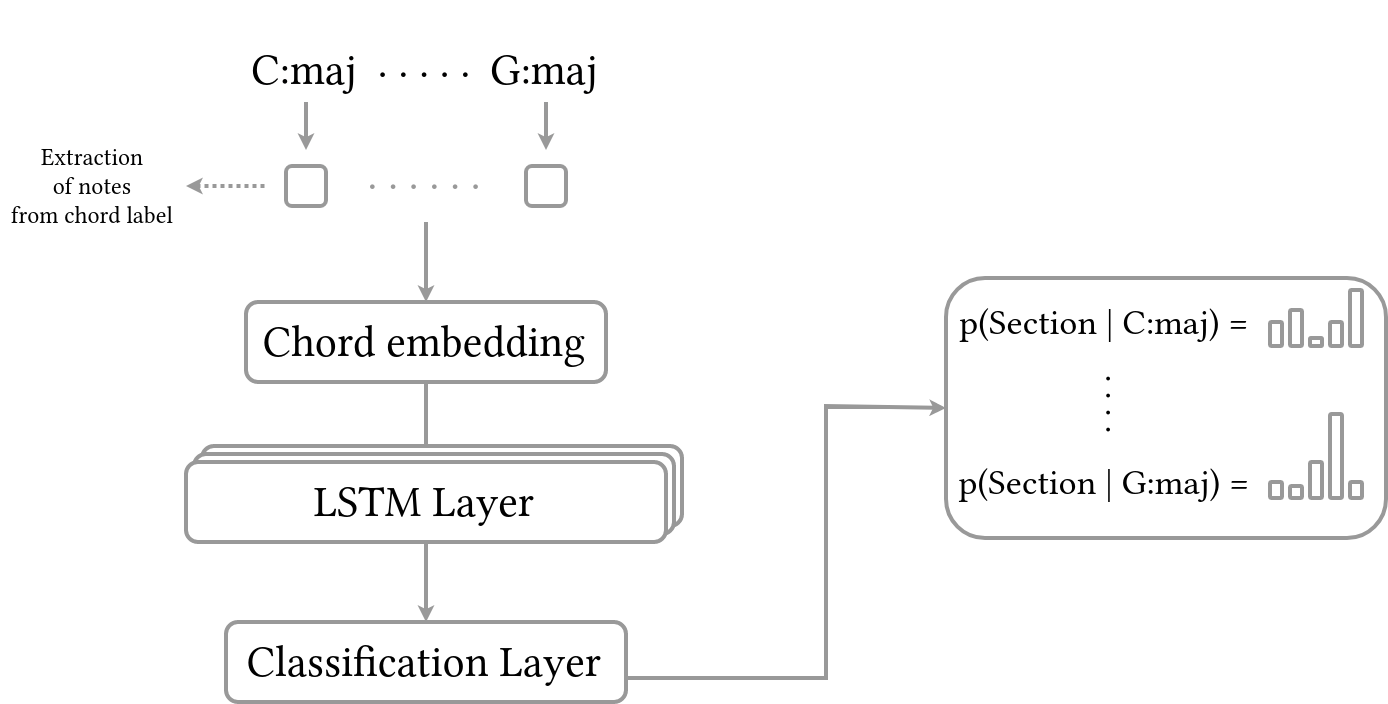}
    \caption{Visual depiction of the implemented LSTM model.}
    \label{fig:lstm-model-diagram}
\end{figure}

This section shows how the proposed model compares to FORM \cite{de2013structural}, the state-of-the-art approach in the field of music segmentation.
We develop a baseline model using a stacked LSTM-based neural network, depicted in figure \ref{fig:lstm-model-diagram}.
The model objective is to predict in which section each chord belongs to. We train our model on the Billboard dataset\cite{burgoyne2011billboard} provided by mirdata\cite{fuentes2021mirdata}. 
The dataset is composed of $889$ expert annotated tracks. Each track is composed of a sequence of chords in Harte format\cite{harte2005symbolic}, and a sequence of structure labels. 
Labels are provided in a similar format to the one presented by SALAMI\cite{smith2011salami}. $80$ unique section labels are present in the whole dataset. We preprocess each label and reduce the number of unique labels to $11$ by combining all those labels that fall under the same definition given by\cite{smith2011salami}.
A complete reference of the label conversion step is given in table \ref{tab:label-conversion}.

\begin{table*}[!ht]
  \caption{Label conversion reference. Each label is stripped out of numbers and symbols before the conversion.}
  \label{tab:label-conversion}
  \begin{tabularx}{\textwidth}{X|l}
    \toprule
    Source labels & Converted label \\
    \midrule
    \texttt{[verse]} & \textit{verse} \\
    \texttt{[prechorus, pre chorus]} & \textit{prechorus} \\
    \texttt{[chorus]} & \textit{chorus} \\
    \texttt{[fadein, fade in, intro]} & \textit{intro} \\
    \texttt{[outro, coda, fadeout, fade-out, ending]} & \textit{outro} \\
    \noindent\parbox[c]{\hsize}{\texttt{[applause, bass, choir, clarinet, drums, flute, harmonica, harpsichord, instrumental, instrumental break, noise, oboe, organ, piano, rap, saxophone, solo, spoken, strings, synth, synthesizer, talking, trumpet, vocal, voice, guitar, saxophone, trumpet]}} & \textit{instrumental} \\
    \texttt{[main theme, theme, secondary theme]} & \textit{theme} \\
    \texttt{[transition, tran]} & \textit{transition} \\
    \texttt{[modulation, key change]} & \textit{other} \\
    \bottomrule
\end{tabularx}
\end{table*}

Although in literature there are neural network architectures that have proven to perform better in similar task \cite{huang2015crfbilstm, shi2015convlstm, wang2016attentionlstm}, we deliberately decided to use a very straightforward architecture.
This is due to the fact that the aim of this study is to compare different types of embedding, rather than to achieve the best performance. 

We split our dataset in the usual training, validation and test split (respectively $800$, $178$ and $89$ elements) and fine-tune each model hyper-parameters (number of LSTM stacked layers, LSTM hidden size, dropout probability) to obtain the best results on the validation set. The final configuration of each model is summarised in Table \ref{tab:best-hyperparameters}. Training is performed using an \textit{NVIDIA RTX 3090} with a batch size of $128$. Each model takes at most few minutes to train and average less than $2$ milion parameters.

\begin{table}[b]
    \caption{Best hyper-parameters obtained on the validation set for each model.}
    \begin{tabularx}{\textwidth} { X|c|c|c }
        \toprule
        Model & Hidden size & Number of stacked layers & Dropout probability \\
        \midrule
        word2vec & 100 & 5 & 0.3 \\
        fasttext & 100 & 5 & 0.5 \\
        pitchclass2vec & 100 & 10 & 0 \\
        pitchclass2vec + word2vec & 200 & 5 & 0.3 \\
        pitchclass2vec + fasttext & 200 & 5 & 0 \\
        \bottomrule
    \end{tabularx}
    \label{tab:best-hyperparameters}
\end{table}

We compare the proposed embedding model to \textit{fasttext} and \textit{word2vec} as well, in which both methods are trained on the string labels of chords in Harte format. Both the models are trained using the highly optimized gensim \cite{rehurek2010lrec} implementation. The hyperparameters used are the same as the one described in section \ref{sec:implementation-details} except for the embedding dimension, which is set to $300$.
\section{Results}
\label{sec:results}
\begin{table}[b]
    \caption{Evaluation metrics on the test set. FORM\textsubscript{raw} is computed on the same chord labels that are used for all the neural approaches. FORM\textsubscript{simple} is computed on simplified chord representation as done in \cite{de2013structural}: only the chord \textit{root} and its quality (\textit{major} or \textit{minor}) are kept. }
    \begin{tabularx}{1.01\textwidth} { X|c|c|c|c|c|c }
        \toprule
        Method & $P$ & $R$ & $F1$ & $S_U$ & $S_O$ & $S_{F1}$ \\
        \midrule
        FORM\textsubscript{raw} & 0.667640 & 0.338019 & 0.425610 & 0.667640 & 0.338019 & 0.425610 \\
        FORM\textsubscript{simple} & \textbf{0.681281} & 0.325479 & 0.416651 & 0.681281 & 0.325479 & 0.416651 \\
        \hline
        word2vec & 0.410190 & 0.823330 & 0.523692 & 0.605202 & 0.257582 & 0.360186 \\
        fasttext & 0.373044 & \textbf{0.993201} & 0.526918 & \textbf{0.947381} & 0.154424 & 0.264553 \\
        pitchclass2vec & 0.402290 & 0.953399 & 0.547694 & 0.719733 & \textbf{0.431959} & \textbf{0.537879} \\
        \noindent\parbox[l]{\hsize}{pitchclass2vec + word2vec} & 0.467940 & 0.664824 & 0.532202 & 0.544820 & 0.415806 &  0.471398 \\
        pitchclass2vec + fasttext & 0.433019 & 0.835007 & \textbf{0.553045} & 0.539774 &  0.425738 & 0.473986 \\
        \bottomrule
    \end{tabularx}
    \label{tab:results}
\end{table}

The results of the experiments are summarised in Table \ref{tab:results}. We evaluate the segmentation results by computing pairwise precision, recall and F1-score ($P$, $R$ and $F1$ in Table \ref{tab:results}) \cite{levy2008structural} along with under-segmentation, over-segmentation and normalized cross entropy F1 ($S_U$, $S_O$ and $S_{F1}$ in table \ref{tab:results}) \cite{lukashevich2008towards}. Every metric is computed using the standard MIR evaluation library mir\_eval \cite{raffel2014mireval}.
\textit{Under-segmentation} and \textit{over-segmentation} are two peculiar metrics for the evaluation of automatic music segmentation methods.
When a method has an high over-segmentation measure, the final prediction accuracy is influenced mostly by false fragmentation. Conversely an high under-segmentation measure means that the prediction's segments are the result of ground-truth segments being merged together \cite{lukashevich2008towards}.

\medskip
\begin{figure}
\centering
\begin{subfigure}{\textwidth}
  \centering
  \includegraphics[width=.7\textwidth]{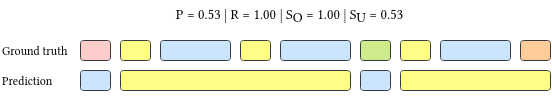}
  \caption{High over-segmentation example. $R = 1$ since if we take each chord in the sequence pairwise then each chord that should be in the same section is indeed in the same section. $S_O = 1$ since the accuracy of the prediction can be easily explained by the over-segmentation phenomena. Conversely, $P = 0.53$ and $S_U = 0.53$ clearly show how the prediction is not able to capture all the needed segments but rather merges ground truth segments together. }
\end{subfigure}
\begin{subfigure}{\textwidth}
  \centering
  \includegraphics[width=.7\textwidth]{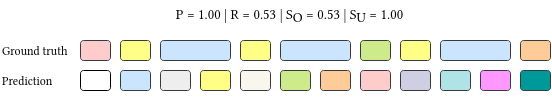}
  \caption{High under-segmentation example. The exact opposite of Figure (a) is displayed. The prediction is not able to capture segments and rather place each chord on its own segment. }
\end{subfigure}
\begin{subfigure}{\textwidth}
  \centering
  \includegraphics[width=.7\textwidth]{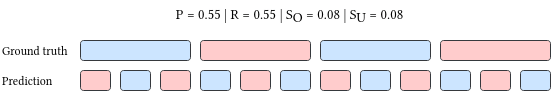}
  \caption{$P$, $R$ and $S_O$, $S_U$ compared. In this edge case the main difference between the two measures is highlighted. While $P$ and $R$ suggests a decent segmentation $S_O$ and $S_U$ clearly states a completely wrong segmentation. Pairwise metrics can be misguiding in absence of $S_O$ and $S_U$. }
\end{subfigure}
\caption{Examples of metric computation on relevant instances.}
\label{fig:measures-examples}
\end{figure}

Pairwise metrics are computed as the usual precision, recall and F1 scores on the set of identically labeled pairs in the sequence.
Precision and recall can be interpreted as the amount of accuracy that is influenced respectively by under-segmentation and over-segmentation.
On the other hand under and over-segmentation scores are computed by taking into account the normalized conditional entropy of the segmentation. 
In short, $S_O$ gives a measure of how much information is missing in the predicted segmentation, given the ground truth segmentation while $S_U$ gives a measure of how much noisy information are the result of the predicted segmentation \cite{lukashevich2008towards}.
A graphical explanation of these concepts is provided in Figure \ref{fig:measures-examples} (all the examples are taken from \cite{lukashevich2008towards}).

We evaluate our models based on the $F1$ and $S_{F1}$ scores of Table \ref{tab:results} since both metrics gives a balanced measure of over and under segmentation.

FORM\textsubscript{simple} detect repetitive patterns from simplified chord labels, as shown in \cite{de2013structural}. The chord simplification process extracts the \textit{root} note from the chord and classifies it either as \textit{major} or \textit{minor}. FORM\textsubscript{raw} uses the same labels used by the neural approaches.
The former performs better better than the latter. This is not surprising as more patterns between strings can be uncovered by only taking into account $24$ labels ($12$ root notes, each of which can be either \textit{minor} or \textit{major}). 
$S_O$ and $S_U$ however suggests that the over-segmentation based approach of FORM ends up correctly segmenting only some particular portions of the whole composition, while the other ones are wrongly classified. This is an expected behaviour since FORM only relies on label-based repeated patterns. 
The presence of subtle differences in an harmonic progressions that belongs to the same section, such as the first and second \textit{verse} in Figure \ref{fig:helterskelter-structure}, are not detected.

All the neural models in table \ref{tab:results} outperforms FORM. Even though each neural model shows some differences in term of metrics, the clear trend is that syntactical-based models (\textit{fasttext} and \textit{word2vec}) yield overly segmented results, as the low $S_O$ score suggests, while the approach taken by \textit{pitchclass2vec} produces a more balanced segmentation, as suggested by the $F1$ score.
Surprisingly, \textit{fasttext} and \textit{word2vec} under-performs when compared to the FORM baseline on $S_{F1}$ metric. The model is not able to generalize enough over the representation and cannot detect patterns that are detected by FORM.
Finally, the combination of \textit{pitchclass2vec} with either \textit{fasttext} or \textit{word2vec} doesn't bring any remarkable benefit to our novel representation. Even though an higher $F1$ score is obtained by using an hybrid approach, the lower $S_{F1}$ score suggests that it has an underlying less accurate segmentation, similar to example (c) in Figure \ref{fig:measures-examples}.
\section{Conclusion and Future Work}\label{sec:conclusion}

In this article, we presented a new embedding method for musical chords, \textit{pitchclass2vec}, that considers the component notes of the chord (also called \textit{pitchclass}), instead of the chord label, as used in embedding methods in the natural language processing field.  
In addition, we proposed hybrid embedding forms, which combine embedding on the chord label and the novel \textit{pitchclass2vec}. 
We compared different embedding models, including \textit{pitchclass2vec} with the state-of-the-art approach in the field of music structural segmentation. 
We used ChoCo, a dataset of chord annotations, for training the embeddings and Billboard, a dataset of structurally annotated tracks, for the music segmentation task. We used the different types of embeddings on a recurrent neural network (LSTM).
The results obtained by using our embeddings outperform the state of the art in every case, with the best result obtained by \textit{pitchclass2vec}, achieving a pairwise F1 score of $0.548$ and an over-under-segmentation F1 score of $0.538$. 

\textit{Pitchclass2vec} is effectively able to learn the harmonic relationships that ties different chords together. Even though the experiments based on \textit{fasttext} and \textit{word2vec} proves to be effective as well, the musical theoretical approach upon which we base \textit{pitchclass2vec} is an essential factor that needs to be taken into account. The presented embedding model proves to be a promising method to improve results in MIR tasks that can be complemented with harmonic information. Moreover, it provides a valuable tool to better understand and analyse harmonic progressions, since it allows a richer comparison between chords and chord sequences when compared to string labels.


There are additional information that we plan to integrate on \textit{pitchclass2vec} to obtain a richer and more accurate representation.
For instance, one of the main limitations of our approach stems from the fact that we do not take temporal information into account. 
We plan to test this possibility by using the temporal information directly in the embedding process and further modify the LSTM model to condition the classification of the section of a chord based on its duration.
As discussed in Section \ref{sec:model}, chord labels have their own semantic as well. 
Since the hybrid models proposed did not directly result in more accurate results, we plan on expanding the \textit{pitchclass2vec} method to take into account the label of a chords as well directly in its embedding model.
To obtain a semantically richer representations we plan on enhancing \textit{pitchclass2vec} by using deep contextual word embeddings \cite{peters2018elmo} along with knowledge enhancement techniques \cite{peters2019knowledge} that combines domain-specific ontologies, such as \cite{kantarelis2023functional}, with deep contextual word embeddings.

Moreover, we plan an in-depth analysis of \textit{pitchclass2vec} training parameters, described in section \ref{sec:implementation-details}, since in \cite{caselles2018word2vec} the authors showed that, on a product recommendation task, carefully optimized hyper-parameters nearly double the final accuracy on all the experiments.

It is worth mentioning that the LSTM model that we implemented for the structure segmentation task does not take advantage of two fundamental aspect of the task itself: the segmentation of a musical piece should be conditioned by its musical genre. 
In fact, the annotation guidelines provided in \cite{smith2011salami} defines some genre-specific labels and encourage their use whenever applicable. 
Even though a relabeling process can partially solve this issue, the need of genre-specific labels proves that the use-cases of specific structure labels, for instance \textit{theme}, might be different in different musical genres.
Finally, as already discussed in section \ref{sec:experiments}, many recent techniques has shown to be effective in increasing the accuracy of different tasks in the NLP field \cite{huang2015crfbilstm, shi2015convlstm, wang2016attentionlstm} when using continuous word representations. We plan to address these issues in future works.


\begin{acknowledgments}
This project has received funding from the European Union’s Horizon 2020 research and innovation programme under grant agreement No 101004746.
\end{acknowledgments}
\bibliography{references}

\appendix

\end{document}